%% file: main.tex
\documentclass[sigconf,10pt]{acmart}
\usepackage{xspace}
\usepackage{subfig}
\usepackage{multirow}
\usepackage[compact]{titlesec}
\usepackage[font=small,labelfont=bf]{caption}

\usepackage{array}
\usepackage{hyperref}
\hypersetup{
    colorlinks=true,
    linkcolor=blue,
    filecolor=blue,
    urlcolor=blue
}

\usepackage{listings}
\usepackage{xcolor}

\definecolor{codegreen}{rgb}{0,0.6,0}
\definecolor{codegray}{rgb}{0.5,0.5,0.5}
\definecolor{codepurple}{rgb}{0.58,0,0.82}
\definecolor{backcolour}{rgb}{0.95,0.95,0.92}

\newcommand{\utaustin}{\affiliation{\institution{The University of Texas at Austin}\city{Austin}\state{TX}\country{USA}}}

\newcounter{packednmbr}

\def\Snospace~{\S{}}

\newcommand{\framework}{\textsc{PolicySmith}\xspace}

\newcommand{\checker}{\textsc{Checker}\xspace}
\newcommand{\evaluator}{\textsc{Evaluator}\xspace}
\newcommand{\generator}{\textsc{Generator}\xspace}
\newcommand{\template}{\textsc{Template}\xspace}

\newcommand{\heuristic}[1]{Heuristic~$\mathcal{#1}$}

\newcommand{\ie}{\textit{i.e., }}
\newcommand{\eg}{\textit{e.g., }}

\title{Man-Made Heuristics Are Dead. Long Live Code Generators!}

\author[Dwivedula, Saxena, Akella, Chaudhuri, and Kim]{Rohit Dwivedula \quad Divyanshu Saxena \quad Aditya Akella \quad Swarat Chaudhuri \quad Daehyeok Kim} \utaustin

\acmYear{2025}\copyrightyear{2025}
\setcopyright{acmlicensed}
\acmConference[HOTNETS '25]{The 24th ACM Workshop on Hot Topics in Networks}{November 17--18, 2025}{College Park, MD, USA}
\acmBooktitle{The 24th ACM Workshop on Hot Topics in Networks (HOTNETS '25), November 17--18, 2025, College Park, MD, USA}
\acmDOI{}
\acmISBN{}

\begin{abstract}
Policy design for various systems controllers has conventionally been a manual process, with domain experts carefully tailoring heuristics for the specific instance in which the policy will be deployed.
In this paper, we re-imagine policy design via a novel automated search technique fueled by recent advances in generative models, specifically Large Language Model (LLM)-driven code generation.
We outline the design and implementation of \framework, a framework that applies LLMs to synthesize instance-optimal heuristics. 
We apply \framework to two long-standing systems policies - web caching and congestion control, highlighting the opportunities unraveled by this LLM-driven heuristic search.
For caching, \framework discovers heuristics that outperform established baselines on standard open-source traces.
For congestion control, we show that \framework can generate safe policies that integrate directly into the Linux kernel.
\end{abstract}

\begin{document}

\maketitle

\input{sections/intro}

\input{sections/heuristic-design}
\input{sections/framework}
\input{sections/cache}
\input{sections/cong_control}

\input{sections/discuss}

\input{sections/related}

\bibliographystyle{ACM-Reference-Format} 
\bibliography{citations}

\end{document}

%% file: sections/intro.tex
\section{Introduction}
\label{sec:intro}

Systems research has long treated policy design as a manual craft. Whether it is congestion control, caching, or scheduling, performance-critical systems rely on heuristics hand-written by domain experts, optimized for typical conditions, and deployed as a static policy. Yet these policies are brittle, degrading under new workloads, hardware, or objectives.

The standard response is to continually tune and re-implement heuristics: we tweak congestion control algorithms for new network environments~\cite{pudica-cc-for-cloud-gaming}, tailor prefetching and caching policies for emerging workloads~\cite{fetchbpf} or hardware~\cite{cachesack-atc22}, and evolve queueing disciplines for new performance targets~\cite{cake}. But this is an arms race we are losing. We face rapid evolution in workloads, deployment settings, and heterogeneous hardware. As such, the design space of heuristics for most problems is complex and shifting in a context-dependent manner. Human developers often cannot discover "the right heuristics" fast enough.

Meanwhile, learning-based systems have shown that it is possible to approach \emph{instance-optimality}, i.e., finding the best policy for a given context by learning from data. Prior work in learned congestion control~\cite{orca} and caching~\cite{lrb,glcache,3l-cache}, for example, demonstrates that data-driven policies, represented as neural networks, can significantly outperform fixed heuristics. However, neural approaches come at a steep cost: opaque behavior~\cite{metis-sigcomm20}, complex training and deployment pipelines~\cite{lake,kml-lib}, inference overheads~\cite{lake,liteflow-sigcomm22}, and safety concerns~\cite{guardrails-hotos} that preclude adoption in many environments.

Alternatively, we find that recent advancements in coding agents such as FunSearch~\cite{funsearch}, AlphaEvolve~\cite{alphaevolve}, and EvoPrompting~\cite{evoprompting} present a powerful new opportunity for policy design. These agents use a form of LLM-guided evolutionary computation to synthesize expressive code that maximizes quantitative reward functions. We propose to use this approach to automatically synthesize system policies.

In our approach, {\em policy design is re-imagined as an automated search problem, solved as often as needed using generative models to produce instance-optimal policy code}. This means that ``intelligence'' in systems no longer needs to reside in hard-coded rules or opaque neural network weights. Instead, it could reside in the {\em process} by which policies are generated. This enables a move away from deploying fixed logic to a pipeline that can automatically generate, evaluate, and produce optimized code for each context. By pairing LLMs with evolutionary search and test-time feedback, we can explore vast policy spaces and discover high-performing code that would be infeasible to author manually. This approach also avoids opaque learned models in favor of policy code that is safer and more interpretable.

In the limit, this shift can unlock a future where systems come with tailor-made, self-evolving policy logic, allowing developers to simply control high-level specifications that include metrics they care about, such as performance and an upper bound on the costs incurred to "search" for the policy, among other things.

\noindent{\bf One instantiation.}
To ground this idea, we sketch \framework, a prototype system that uses LLMs to synthesize instance-optimal policies. Given an objective and test harness, it generates, evaluates, and evolves policy code offline. However, our broader contribution is not a tool---it is a call to rethink the boundary between systems and machine learning, and to embrace a new model of policy design.

%% file: sections/heuristic-design.tex
\section{Heuristic Design: Status Quo and Vision}
\label{sec:example-with-caching}
Systems research has seen a long history of heuristics, such as cache eviction, congestion control, and queue scheduling, being developed, modified, and fine-tuned for a specific ``context'' which is defined by the workload (application and traces) being supported by the heuristic, the desired objectives (\eg performance, utilization, fairness, scalability, etc.), and the environment (\eg hardware) where the heuristic is running.

Taking the concrete example of web caching, different eviction heuristics for specific traces, objectives, or deployment scenarios~\cite{twoq, lhd, arc, sieve, halp, s3-fifo, cacheus, lecar}.
ARC~\cite{arc} and SIEVE~\cite{sieve} perform well for large cache sizes by balancing new and old objects.
In contrast, TwoQ~\cite{twoq} and LHD~\cite{lhd} perform well for smaller caches due to their ability to quickly remove low-value objects.
Cacheus~\cite{cacheus} shows that depending on whether a workload consists of mostly new objects (``scan workloads'') or mostly repeated objects (``churn workloads''), different expert algorithms perform better. 
Additionally, heuristic design often takes into account end-to-end \textit{objectives} such as tail latency~\cite{robinhood-osdi18} and fairness~\cite{robus-sigmod17}, or system-level \textit{constraints} such as CPU overhead~\cite{halp}, lock-free design~\cite{s3-fifo}, and memory efficiency~\cite{tinyLFU}.

Because no single heuristic performs well across all contexts, experts routinely invest significant time adapting or inventing new heuristics for new workloads, objectives, and environments. This process---designing new algorithms~\cite{sieve, s3-fifo, twoq, lhd}, engineering feature sets for learned models~\cite{halp, glcache, lecar}, and evaluating across diverse traces~\cite{halp} using simulators~\cite{libcachesim} or real-world systems~\cite{cachelib}---is manual and painstakingly slow.
This problem is not limited to caching alone; it is widespread across several systems policies. As a result, complex systems (\eg the Linux Kernel) often continue running suboptimal policies for years, because discovering better ones requires developers with kernel expertise and is too labor-intensive.
The ability to automatically discover context-specific, incrementally better heuristics can yield meaningful gains in terms of both performance metrics of interest and manual effort involved. %

\textbf{Our vision} is to automate the \textit{heuristic design process} so humans can focus on what matters most: defining \textit{what} they want. This includes both the \textit{intent}---the goals/objectives of the heuristic---and the \textit{space} it operates in, shaped by context and constraints. These choices require high-level reasoning about system goals, tradeoffs (like fairness or liveness), and operator needs---which are difficult to formalize and automate, and deeply benefit from human experience and intuition. Once the intent and space are clear, generating heuristics becomes a \textit{search} problem that can be scaled, guided by feedback, and automated.

In particular, automating this search process is feasible today because large language models (LLMs) are remarkably effective \textit{generators}: they can quickly produce a wide range of candidate heuristics, remixing and adapting known techniques across domains. Many state-of-the-art heuristics, such as SIEVE~\cite{sieve}, ARC~\cite{arc}, and Cacheus~\cite{cacheus}, are delicate recombinations and improvements of existing approaches. Because LLMs have been pretrained on code and patterns from across the stack, they are well-positioned to generate {\em candidate} heuristics inspired by these recurring structures; though inventing entirely new structures and principles may still require human insight.

%% file: sections/framework.tex
\begin{figure}[t!]
    \centering
    \includegraphics[width=0.95\linewidth]{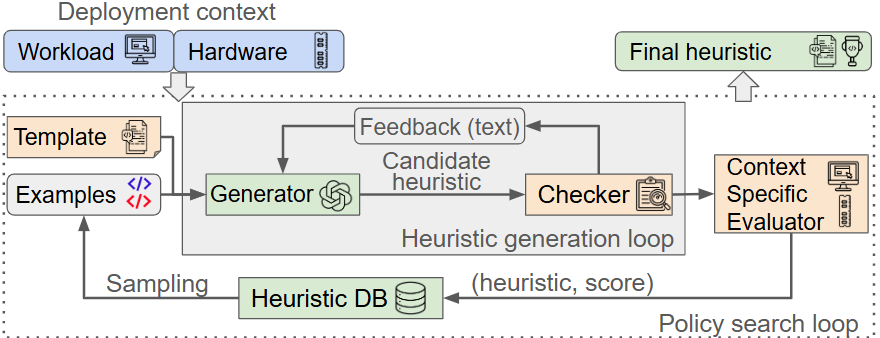}
    \caption{\framework overview}
    \Description{Diagram of a \framework's policy search mechanism. The framework takes as input a deployment context (workload, hardware) and produces a final heuristic tailored for this context. It shows an internal loop where a Generator uses templates and examples to create candidate heuristics, which are checked by a Checker - in case of any issues, the checker provides structured feedback in natural language for the generator to refine the heuristic. Once the candidate heuristic passes the checker's checks, it is evaluated by a Context-Specific Evaluator, and a Heuristic DB stores heuristics and scores, enabling sampling.}
    \label{fig:framework-overview}
\end{figure}

\section{\framework}
\label{sec:framework}
To turn the vision of automated heuristic design into practice, we build the \framework framework (Fig~\ref{fig:framework-overview}) that separates specification (user-defined) from policy search (automated). 

The user is responsible for designing a \template, which defines the space of programs to search, and an \evaluator that runs candidate heuristics and returns a numeric \textit{score} indicative of how well it performed in the current deployment context. The \template contains a function signature or a partial code stub that specifies the semantics of what the heuristic must implement.
For example, in a caching system, this might be a function that decides which object to evict from a set of eviction candidates; for congestion control, it could be event-driven functions that adjust the \textit{cwnd} on packet ACK or packet loss.

The \template, used by an LLM-based \generator to produce candidate heuristics, also contains natural-language \textit{constraints} that guide generation. 
These constraints describe the constructs allowed in the heuristic, including which libraries may be imported, the states it can access, and any behavioral or performance requirements. 
For example, congestion control heuristics on the critical path in the kernel must avoid floating-point arithmetic, locking, or unbounded loops. In caching, constraints may require \texttt{O(log N)} complexity, ruling out full-cache scans.
Overall, the \template (including the constraints) acts like a "design spec", ensuring generated heuristics are efficient and deployable.

The LLM, of course, may produce code that does not honor these constraints, due to hallucination~\cite{llm-hallucinations-survey,deepmind-delusions-in-lms}, producing plausible yet non-conforming or incorrect code.
To catch such violations, users define a \checker that enforces syntactic and semantic rules, and provides structured feedback to help generators stay within spec.

With these components defined, the framework begins its search loop. 
Motivated by recent coding agents such as AlphaEvolve \cite{alphaevolve} and FunSearch \cite{funsearch}, we leverage evolutionary search to explore the large space of programs defined by the \template.  This choice is motivated by the fact that the space of ``good" heuristics is discrete, combinatorially large, and sparsely populated -- making gradient-based methods impractical and local search~\cite{stochastic-superoptimization} limited in efficacy.

In our search process (Fig~\ref{fig:framework-overview}), an LLM-based \generator synthesizes multiple candidate heuristics based on the \template, which are scored by the context-specific \evaluator, and the best-performing candidates are fed back as examples in the next round. 
This loop continues for several iterations, gradually steering the generator toward better-performing heuristic code. At the end, \framework outputs a final heuristic tailored to the target context.

\subsection{Responding to Context Shifts}
\label{sec:implicit_explicit_context}
\framework generates instance-optimal heuristics for each \textit{context}, defined as a combination of the workload, hardware, and the desired objectives. However, how is a context actually delineated in practice? How do we identify when the deployment context has drifted enough to warrant re-synthesis?

\subsubsection{Explicit context shifts.} Some context changes might be obvious. For instance, upgrading from HDDs to SSDs usually requires a new block I/O scheduling heuristic~\cite{flashfq} even if nothing else changes (hardware change); different application classes on the same hardware benefit from different process scheduling~\cite{ghost} and networking queuing heuristics~\cite{no-silver-bullet-data-plane} (workload change).
In such scenarios, \framework can be invoked manually for each context. 

\subsubsection{Implicit context shifts.} 
Context changes are not always readily apparent. A CDN server, for instance, may experience shifts in access patterns due to the time of day, even if the hardware and objectives remain unchanged. A cloud provider may not have visibility into application-level changes that affect the performance of heuristics. These drifts degrade the performance of previously specialized heuristics without triggering any explicit event.

To address this, prior work has developed runtime adaptation systems that leverage online techniques such as reinforcement learning~\cite{cdbtune-sigmod19}, bayesian optimization~\cite{cherrypick}, and multi-armed bandits~\cite{configanator,darwin} to identify context changes and respond to them, by selecting a new heuristic or configuration from a pre-defined set. These systems rely on lightweight monitoring infrastructure (\eg guardrails~\cite{guardrails-hotos}) or clustering-based approaches~\cite{darwin} to detect context shifts.

\framework complements these approaches: the same monitoring signals that guide these systems can also trigger \framework automatically, allowing it to re-synthesize a heuristic offline.
Until a new, better heuristic is ready, the existing one continues to be used with degraded performance. Over time, this enables building a \textit{library} of \framework-generated heuristics, providing better options for an adaptation system to choose from.

This paper does not focus on designing context-detection or runtime-adaptation systems, and rather assumes such triggers (manual or automated) are available. Given such a trigger, the core question then becomes: can we reliably synthesize effective heuristics for the new context? We will explore this in two case studies\footnote{All code used for these case studies is available at \href{https://github.com/ldos-project/policysmith}{https://github.com/ldos-project/policysmith}}.

First (\S\ref{sec:caching_concrete_details}), we use \framework to discover instance-optimal cache eviction heuristics, illustrating its ability to generate high-performing, context-specific policies. In the second case study (\S\ref{sec:congestion}), we use \framework to evolve Linux kernel code to synthesize congestion control heuristics, demonstrating the feasibility of our vision even in highly constrained, safety-critical systems such as the Linux kernel.

%% file: sections/cache.tex
\section{Case Study: Web Caching}
\label{sec:caching_concrete_details}

We now describe an instantiation of \framework to find instance-optimal eviction policies for web caches.
Our prototype is built on libCacheSim~\cite{libcachesim}, a high-performance web cache simulator with an event-driven interface.
We describe our prototype design (\S\ref{sec:cache-framework-design}), followed by results (\S\ref{sec:cache_results}).

\subsection{Design}
\label{sec:cache-framework-design}
\subsubsection{Tradeoffs in template design.}
Designing an effective \template for heuristic generation in \framework requires deciding which parts of the heuristic should be exposed to the \generator and which to hold fixed. 
Many heuristics (including web caching) naturally consist of two parts: \textit{state management} (e.g., tracking metadata) and \textit{decision-making logic} (e.g., choosing what to evict). 

One option is to use the \generator to synthesize both. However, this requires the LLM to coordinate logic and state across multiple interdependent functions. This increases the size of each candidate heuristic, increases the computational cost, and is more likely to result in errors due to complexity. 
At the other extreme, we could fix the state management---\ie the data structures used to track cache metadata---to match a known policy (\eg GDSF~\cite{gdsf}, ARC~\cite{arc}) and evolve only the decision logic.
While this simplifies code generation, it severely limits what can be discovered: it becomes impossible to discover heuristics that depend on richer signals like access history, temporal patterns, or global cache statistics.

\begin{table}[t]
\centering
\small
\begin{tabular}{p{1.6cm}p{5.7cm}}
\toprule
\textbf{Type} & \textbf{Attributes} \\
\midrule
Per object & Number of accesses (count), last access time, time added to cache, object size \\
Aggregates & Percentiles over access counts, ages, or sizes (\eg p50 size in bytes of all objects in cache).  \\
History & List of recently evicted objects, along with (timestamp, access count, age) at eviction. \\
\bottomrule
\end{tabular}
\caption{Features available to \texttt{priority()}.}
\label{tab:cache-features}
\vspace{-0.2in}
\end{table}

\subsubsection{Template definition.}\label{subsubec:cache-template}
For this case study, we use a \template that favors simplicity and ease of generation.
In our design, object metadata is stored in a priority queue, with the position determined by a customizable \texttt{priority()} function, that is synthesized by the \generator. This \texttt{priority} function is invoked on each access or insertion and updates the object's priority score; when needed, the object with the lowest score is evicted. To support diverse strategies, \verb|priority()| has access to a rich set of features (Table~\ref{tab:cache-features}) designed as a superset of the features used by existing caching policies. 
This enables the discovery of varied heuristics without changing the underlying queue structure or interface.

The resulting heuristic may incur higher overhead than policies like LRU or FIFO, due to \texttt{O(log N)} priority updates on each access.
This may be acceptable for caches with relatively fewer objects (our focus in this case study).
For caches where the overhead is prohibitive, alternative \template designs with stricter constraints, such as using approximate structures like soft heaps~\cite{soft-heap}, may be needed.

\subsubsection{Generator.}
Because our \template is narrow and self-contained, we use a lightweight LLM, GPT-4o mini~\cite{gpt-4o-system-card} via the OpenAI API, for heuristic generation. Using a smaller model like this keeps generation fast and inexpensive; more broadly, the complexity of the \template governs the sophistication of the required \generator. As templates become richer or span multiple functions, they may require larger, more capable models to handle reasoning and coordination. As LLMs continue to improve, we expect this design space to widen,  enabling the evolution of heuristics in more complex templates.
The narrow \template implies that most errors surface as build failures, allowing for a simple \checker that automatically feeds errors back to the \generator. 

\subsubsection{Context-specific evaluator.}\label{subsubsec:cache-evaluator} 
In caching, the \textit{context} is defined by the cache size (typically constrained by hardware) and the \textit{access patterns of requests} (\ie the workload). For our case study, we use two real-world block I/O trace datasets: CloudPhysics~\cite{cloud-physics-dataset} (with 105 week-long traces from diverse VMs) and Microsoft Research Cambridge (MSR) dataset~\cite{msr-dataset} (with 14 traces from production servers). 
In our case study, the \evaluator scores candidate heuristics by running them on \textit{a single} block I/O trace from these datasets, for a fixed cache size (10\% of the trace footprint).
Each pair (trace, cache size) defines a distinct \textit{context}, and the objective is to minimize the object miss rate. While we define context narrowly for simplicity, one could alternatively treat an entire distribution of traces---\eg a sample from all 105 CloudPhysics traces---as a single, broader context, evaluating heuristics based on an aggregate metric such as average hit rate.

\subsection{Results}
\label{sec:cache_results}
\subsubsection{Methodology.} 
\label{subsubsec:caching_reults_methodology}
We select \textbf{one} CloudPhysics trace (\verb|w89|) and use it as the \textit{context}. 
The prompt to the \generator includes: a natural language description of our priority queue interface and available features (Table~\ref{tab:cache-features}), the function signature for \texttt{priority()}, and example priority functions seeded at the start of the search---namely, for LRU and LFU.
In each round of \framework, we prompt the \generator repeatedly to generate 25 candidate heuristics, which are then evaluated by the \evaluator.
The top two performing heuristics across all previous rounds are then used as examples in the next round. This process is repeated 20 times, yielding a total of 500 heuristics. The best-performing heuristic from this search, referred to as \heuristic{A}, is shown in Listing~\ref{lst:priority}.
We repeat this process independently for three more CloudPhysics traces to obtain heuristics $\mathcal{B}$, $\mathcal{C}$, and $\mathcal{D}$, and on four MSR traces to obtain heuristics $\mathcal{W}$, $\mathcal{X}$, $\mathcal{Y}$ and $\mathcal{Z}$.

\subsubsection{Baselines and Metrics.}
\label{sec:caching-baselines}
We use fourteen eviction algorithms as our baselines: GDSF~\cite{gdsf}, S3-FIFO~\cite{s3-fifo}, SIEVE~\cite{sieve}, LIRS~\cite{lirs}, LHD~\cite{lhd}, Cacheus~\cite{cacheus}, FIFO-Reinsertion~\cite{fifo-reinsertion} (denoted FIFO-Re), LeCaR~\cite{lecar}, SR-LFU~\cite{cacheus}, CR-LRU~\cite{cacheus}, LRU, MRU, and FIFO. For each caching heuristic, we report the improvement in the miss ratios over a fixed baseline (FIFO), similar to how~\cite{s3-fifo, sieve} report their results.

\begin{figure*}[t]
    \vspace*{-1em}
    \begin{minipage}[b]{0.68\textwidth}
        \centering
        \subfloat[CloudPhysics Traces]{
            \includegraphics[width=0.48\linewidth]{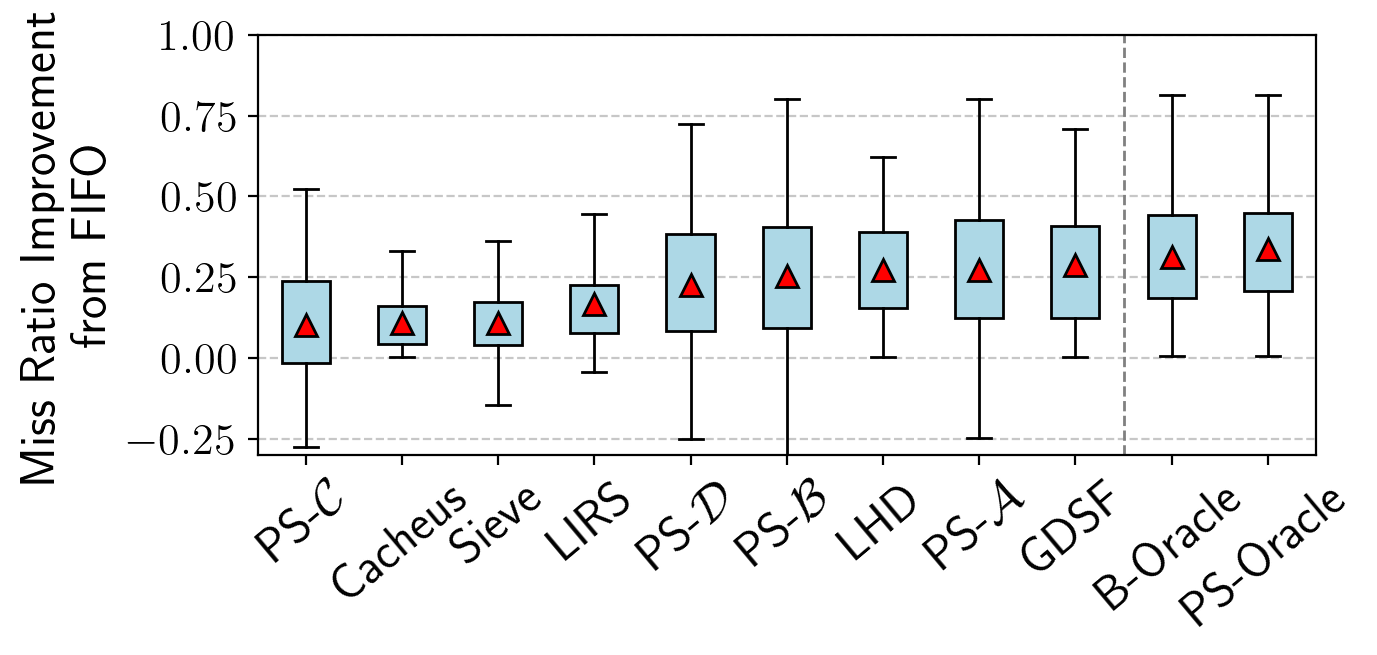}
            \label{fig:cloudphysics-improvement}
        }%
        \subfloat[MSR Traces]{
            \includegraphics[width=0.48\linewidth]{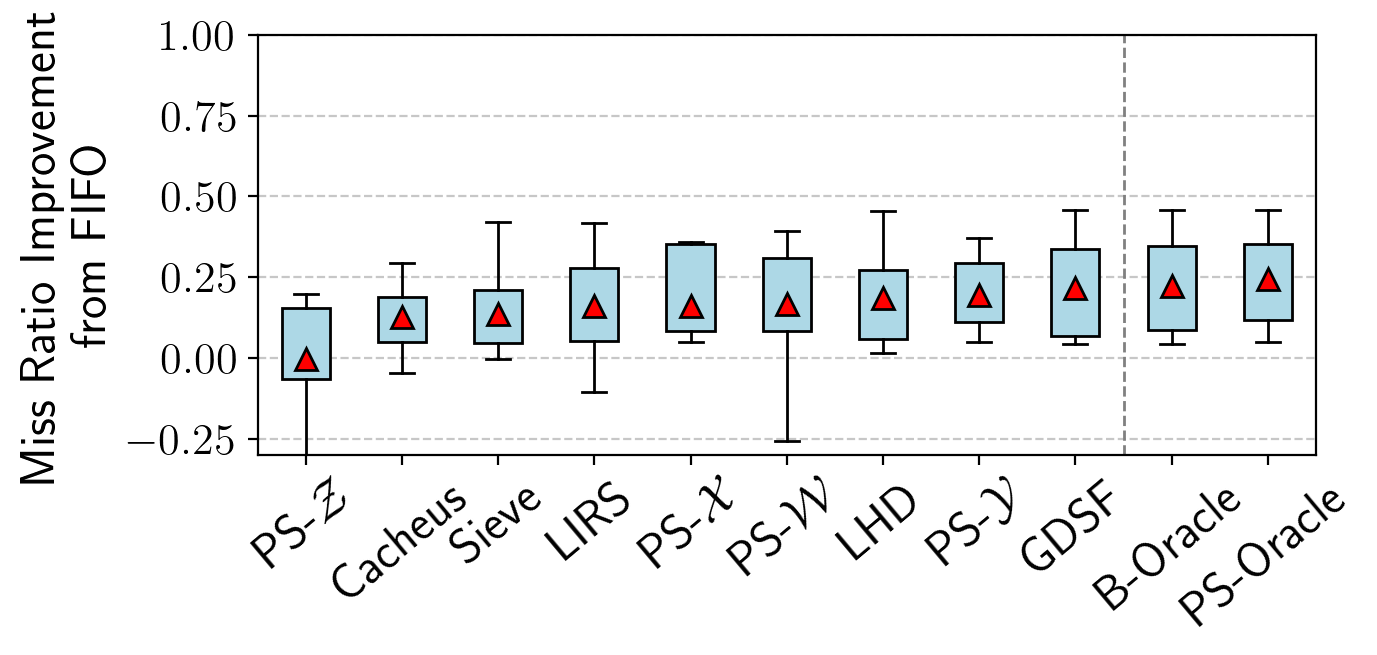}
            \label{fig:msr-improvement}
        }
        \vspace{-1em}
        \caption{Miss ratio improvements over FIFO for all traces in a dataset (higher is better, triangles indicate mean, heuristics ordered left to right by increasing average). Top five performing baselines shown in plot for brevity.}
        \label{fig:caching-improvement}
    \end{minipage}%
    \hfill
    \begin{minipage}[b]{0.3\textwidth}
        \centering
        \small
        \begin{tabular}{lll}
            \toprule
            \textbf{Dataset} & \multicolumn{2}{c}{\%age of traces} \\
            \midrule
            \multirow{2}{*}{CloudPhysics} 
                & $\mathcal{A}$ (48\%) & $\mathcal{B}$ (42\%) \\
                & $\mathcal{C}$ (14\%) & $\mathcal{D}$ (31\%) \\
            \midrule
            \multirow{2}{*}{MSR} 
                & $\mathcal{W}$ (57\%) & $\mathcal{X}$ (64\%) \\
                & $\mathcal{Y}$ (57\%) & $\mathcal{Z}$ (21\%) \\
            \bottomrule
        \end{tabular}
        \captionof{table}{Performance of discovered heuristics: proportion of traces in the dataset where synthesized heuristic outperforms all baselines.}
        \label{tab:generalization}
        \vspace{-1.2em}
    \end{minipage}
    \Description{Box plots of miss ratio improvements for CloudPhysics and MSR traces. For CloudPhysics, heuristics from left to right are: PS-C, Cacheus, Sieve, LIRS, PS-D, PS-B, LHD, PS-A, GDSF, B-Oracle and PS-Oracle. For MSR: PS-Z, Cacheus,Sieve, LIRS, PS-X, PS-W, LHD, PS-Y, GDSF, B-Oracle, PS-Oracle}
\end{figure*}

\subsubsection{Instance-optimality of synthesized heuristics.}
All heuristics produced by \framework ($\mathcal{A}$--$\mathcal{D}$ and $\mathcal{W}$--$\mathcal{Z}$) either match or outperform \textbf{all} 14 baselines for their original context (\ie the trace used in that instance's \evaluator), demonstrating that \framework can tailor heuristics for each context. Additionally, we evaluate each \framework-generated heuristic not just on its original trace, but across all other traces within the same dataset. 
Table~\ref{tab:generalization} summarizes this: for instance, \heuristic{A}, generated using trace \verb|w89|, outperforms all baselines on 48\% of CloudPhysics traces; similarly, \heuristic{X} does so for 64\% of MSR traces. 
This suggests that traces within a dataset may share structural similarities (\eg request access patterns), allowing heuristics tuned on one to remain competitive on others.

While generalization is not our goal, future instantiations of \framework could also define context more broadly, \eg by using multiple similar traces in the \evaluator. Our findings also indicate that the discovered heuristics do not overfit to a single trace.

\input{code/webcache_05}

\subsubsection{Performance of \framework.}
Figure~\ref{fig:caching-improvement} shows the performance of baselines and the \framework-synthesized heuristics on all traces from CloudPhysics and MSR datasets.
Notably, \heuristic{A} and \heuristic{Y} achieve the second-highest average performance (surpassed only by GDSF) for CloudPhysics and MSR datasets, respectively.

The figures also include two \textit{oracles}: (1) the baseline oracle (B-Oracle), which selects the best-performing baseline (\S\ref{sec:caching-baselines}) for each trace, and (2) the \framework-oracle (PS-Oracle), which selects the best heuristic from both the baselines and \framework-synthesized heuristics.
These oracles model ideal runtime adaptation (\S\ref{sec:implicit_explicit_context}), perfectly identifying context and selecting the best heuristic for each trace. They serve as \textit{upper bounds} on expert performance.
The PS-Oracle has a 2\% higher improvement over FIFO than the baseline oracle, demonstrating the additional performance gains possible with the addition of \framework-generated heuristics.

\subsubsection{Example of evolved heuristic.}
Listing~\ref{lst:priority} shows \heuristic{A}, the best performing heuristic we found for CloudPhysics. All of the code in the block, except the function prototype, was generated completely by the LLM. We see that the LLM uses the features provided (Table~\ref{tab:cache-features}) in interesting ways: such as penalizing objects that are old (lines 4, 13) or big (lines 5, 15) and preserving small objects (line 16) or frequent (lines 8, 18).
As discussed in \S\ref{subsubsec:caching_reults_methodology}, the initial seed heuristics provided to \framework are simple algorithms---LRU and LFU---that can be implemented in the priority function in a single line, \eg by returning \verb|obj_info.count| and \verb|obj_info.last_accessed| for LFU and LRU, respectively. 

\subsubsection{Computational cost.}
The search for heuristic~$\mathcal{A}$ required 5.5 CPU-hours to evaluate all candidate heuristics. It also used 800k input tokens and 300k output tokens with the GPT-4o-mini model. The total cost of the OpenAI API for the eight runs in this section was approximately USD \$7.

%% file: code/webcache_05.tex
\lstdefinestyle{mystyle}{
    backgroundcolor=\color{backcolour},
    keywordstyle=\color{magenta},
    emph=[1]{now,obj_id,obj_info,counts,ages,sizes,history},
    emphstyle=[1]\color{teal!70!black},
    emph=[2]{score},
    emphstyle=[2]\color{black}\bfseries,
    numberstyle=\tiny\color{codegray},
    commentstyle=\color{codegreen},
    stringstyle=\color{codepurple},
    basicstyle=\linespread{0.8}\fontsize{8pt}{9pt}\selectfont\ttfamily,
    breakatwhitespace=false,
    breaklines=true,         
    captionpos=b,
    keepspaces=true,                 
    numbers=left,                    
    numbersep=5pt,                  
    showspaces=false,                
    showstringspaces=false,
    showtabs=false,                  
    tabsize=2,
    lineskip=-0.9pt,
    linewidth=0.95\linewidth
}
\lstset{style=mystyle}

\begin{figure}[b]
\begin{minipage}{\columnwidth}
\small
\begin{lstlisting}[style=mystyle,language=C++,
caption={A heuristic discovered by PolicySmith.},
label={lst:priority},frame=single]
priority(now, obj_id, obj_info, counts, ages, sizes, history):
  score = obj_info.count * 20;
  age   = now - obj_info.last_accessed;
  score -= age/300;
  score -= obj_info.size/500;
  if (history.contains(obj_id)){
    h = history.get_metadata(obj_id);
    score += h->count*15;
    score += h->age_at_eviction_time/150;
  } 
  else score-=40;
  recent = ages.percentile(0.75);
  if (obj_info.last_accessed<recent) score-=30;
  big = sizes.percentile(0.75);
  if (obj_info.size > big) score-=25;
  else score+=10;
  frequent = counts.percentile(0.7);
  score += (obj_info.count>frequent) ? 50:-5;    
  if (age < 1000) score+=25;
  if (obj_info.count < 3) score -= 15;
  return score;
\end{lstlisting}
\end{minipage}
\end{figure}

%% file: sections/cong_control.tex
\section{Case Study: Congestion Control}
\label{sec:congestion}
In this section, we explore whether \framework can be extended to a more demanding setting: evolving heuristics in the Linux kernel.
Modern kernels house several policies like TCP congestion control~\cite{cubic,bbr,copa,c2tcp}, packet scheduling~\cite{qdisc-stratified-rr,no-silver-bullet-data-plane,qdisc-stfq,qdisc-scrr}, and block I/O scheduling~\cite{flashfq,kyber,multiqueue-fair}, that have been shaped by decades of manual tuning for specific contexts. 

Instantiating \framework to perform policy search in the kernel is particularly challenging for two core reasons. 
First, the kernel programming environment is highly constrained (\eg floating-point ops disallowed, idiosyncratic patterns to access registers, BPF maps, etc).
These constraints make code generation especially difficult for the \generator{}s, which struggle to produce syntactically and semantically valid code in such narrow, domain-specific contexts~\cite{llms-dsl}. Second, kernel development comes with strict safety and performance requirements: bugs can lead to kernel panics, and even minor inefficiencies can degrade system-wide behavior. This makes the design of the \checker and \evaluator critical.
To evaluate whether these challenges can be overcome, we instantiate \framework in the context of TCP congestion control. This case study is not aimed as a search for new instance-optimal algorithms, but as a focused case study to test whether \framework can navigate the constraints and risks of kernel-space policy generation.

\subsubsection{\template design.}
The Linux kernel requires congestion control algorithms to implement five event-driven callbacks that update the congestion window (cwnd) in response to packet-level events.
Following the \template design used for caching (\S\ref{subsubec:cache-template}), we isolate decision logic from state management, exposing only the decision making function (\verb|cong_control|) to the \generator. This function is provided with a rich set of features, such as previous cwnd, minimum RTT, inflight bytes, among others. To enable reasoning over history, our \template also provides \textit{history arrays} - time series arrays that capture smoothed versions of these metrics over the last 10 RTT intervals~\cite{sage}.

\subsubsection{\template and \checker.}
We implement our \template as a Linux kernel module. However, rather than compiling LLM-generated code directly into the kernel, we offload the generated logic to a dynamically attached eBPF probe that is attached to the \verb|cong_control| Linux kernel function.  At runtime, whenever this function is invoked, the probe is triggered as well.  The eBPF program executes the generated logic, computes the updated decision (\ie cwnd), and writes the result to a BPF map. This design ensures that all candidate programs pass the eBPF verifier~\cite{ebpf} before execution - which acts as the \checker in our framework. This pattern -- using a kernel module for scaffolding and template logic, combined with an eBPF probe for executing generated code and communicating via a BPF map -- is a general mechanism that can be applied to other kernel components such as packet schedulers, I/O controllers, or memory managers.

\subsubsection{Preliminary Results.} We generated 100 candidate congestion control heuristics and attempted to compile them into eBPF programs. Only 63\% of the candidates passed the eBPF verifier on the first try, and an additional 19\% successfully compiled after the \generator was provided with the \verb|stderr|. %
The most common causes of errors were the use of floating-point arithmetic and missing checks for division by zero. This compilation rate for kernel code is substantially lower than what we observed for caching: where 92\% of candidates compiled in the first pass itself.

We evaluated the heuristics that compiled successfully on a 12 Mbps, 20ms delay emulated link~\cite{mahimahi}. The resulting behaviors varied widely: bandwidth utilizations ranged from 23\% to 98\%, and average queuing delays spanned from 2ms to 40ms. This variance illustrates the diversity of policies that can be explored using automated strategies like \framework.

%% file: sections/discuss.tex
\section{Discussion}

\textit{Per-Instance Specialization in Practice.}
Our thesis explicitly abandons the pursuit of ``universal'' heuristics. Instead, we target \emph{instance-optimality}---generating code tailored to each workload, hardware target, and objective function. This specialization raises new challenges: How do we reliably detect when the instance has changed? Can we incrementally update policies? What abstractions enable policy reuse across similar contexts? These are tantalizing open questions, with only a few having initial answers (e.g., guardrails~\cite{guardrails-hotos}).

\textit{Per-Policy vs. End-to-End Coordination.}
Policy synthesis today often targets individual components (e.g., a congestion control algorithm or cache eviction strategy). But in real systems, these policies interact---coordinating across the network stack, storage hierarchy, and compute resources. A key research direction is to extend synthesis to reason about \emph{interactions between synthesized policies}, or even synthesize policies end-to-end across components to achieve global objectives. Can we express cross-cutting goals and constraints? Can LLMs or other program synthesis tools understand and generate such coordinated logic?

\textit{Evaluation Without Deployment.}
Offline policy synthesis hinges on the ability to evaluate candidate policies without full deployment. This raises questions about test harness fidelity, simulation accuracy, and robustness to distribution shift. For safety- or latency-critical systems, policy evaluation must ensure that synthesized logic is not just performant, but correct. Integrating synthesis with formal methods, fuzzing, or worst-case scenario testing may help bridge this gap.

\textit{Trade-offs between Expressivity and Interpretability.}
Models like transformers can capture complex feature interactions via attention, enabling behaviors beyond traditional human-designed heuristics, which are often shallow and simplified. LLM-generated code may offer a middle ground -- more expressive than typical handwritten logic, yet still interpretable and efficient. Moreover, LLMs can be tuned to produce simpler code, preserving interpretability when needed.

\textit{Tools, Workflow, and Developer Experience.}
In this new paradigm, the “programmer” becomes a supervisor of synthesis, not the author of logic. This shift requires a rethink of developer tools: how to prompt, debug, validate, and evolve synthesized policies. %
Understanding how systems engineers interact with these tools---and how to guide synthesis with expert insight---is a key opportunity.

%% file: sections/related.tex
\section{Related Work}

Prior work, such as ~\cite{no-silver-bullet-data-plane,pifo,universal-packet-scheduling}, have hinted at the lack of a universal heuristic and that instance-optimal heuristics are needed for various domains. 
While these efforts propose programmable interfaces, they still rely on developers to craft optimal policies. In contrast, \framework automatically discovers effective heuristics given such an interface. 

Other approaches use solvers~\cite{cat,ccmatic} or program search~\cite{superoptimizer-1987} to generate code, but require detailed system models.
\framework avoids these challenges by using high-level specifications, relying on \generator and \evaluator to search without needing a formal model.

There is a substantial body of recent work on coding agents that combine LLM queries and evolutionary search. For example, FunSearch~\cite{funsearch} uses such an approach for automated discovery of mathematically interesting artifacts; LaSR \cite{lasr} and LLMSR \cite{llmsr} use the approach for scientific discovery; EvoPrompting uses it for neural architecture search; and AlphaEvolve~\cite{alphaevolve} uses it for a range of tasks from mathematical discovery to the synthesis of scheduling heuristics. Our proposed approach is the first to generalize these methods into a unified approach to system policies.

In the context of systems policies, prior work has explored the use of LLMs to automate specific pieces of policy design: such as feature engineering~\cite{nada}, explaining heuristic behavior~\cite{llm-explainability-hotnets24,networking-with-foundation-models-hotnets22}, or synthesizing router configurations~\cite{llm-synthesize-router-config-hotnets22}.

\section{Acknowledgements}
This work was supported by the U.S. National Science Foundation (NSF) Grant Numbers 2326576 and 2212559. Dwivedula was supported with an Amazon AI Fellowship.